# Accelerator Vacuum Windows: A Review of Past Research and a Strategy for the Development of a New Design for Improved Safety and Longevity for Particle Accelerators

C.R. Ader*, M. Alvarez, J.S. Batko, R. Campos, M.W. McGee, and A. Watts
Fermi National Accelerator Laboratory, Batavia, IL 60510, USA

*Abstract*

Vacuum window research continues at Fermilab and this paper will examine cost effective, consistent designs which can have a huge impact on accelerator laboratories in terms of safety and cost.

Issues such as the design, materials, analysis, testing and fabrication are addressed, including beam scattering plots and materials cost-benefit analysis and examining different materials which can potentially be substitutes for beryllium. A previous research paper has examined current fabrication and design techniques and also failure modes at Fermi, and this paper examines previous research in addition to emerging technologies.

Many different paths have been taken by HEP Laboratories throughout the world with varying success. The history of vacuum window development is extensive and not well defined, and a matrix of what research has already been done on materials and joint design for vacuum windows will be shown.

This report finally includes a treatise for vacuum window technology and a view towards emerging designs and materials and discusses future advances of research such as fabrication techniques including additive manufacturing and ultrasonic welding. Further exploration into these would prove beneficial to developing vacuum windows that are safer and stronger while being more transparent to the beam.

## INTRODUCTION

There are approximately 83 vacuum windows in operation at Fermi National Accelerator Laboratory, five of which made of beryllium. Beryllium is typically used in Target halls because of its thermal properties and very low Z-properties. However, if a beryllium window fails, it contaminates the beam-line, and potentially the entire beam enclosure because it is toxic. A window that would be designed to replace beryllium would not contaminate a beam-line if it does fail.

Vacuum windows are a critical part of the beam-line but are also the most fragile component of a vacuum system. They are typically installed upstream of a target, abort dump, or beam stop; they are also used to separate vacuum sectors, or two are installed in a beam-line with a gap between where instrumentation can be installed.

Thin vacuum windows have been used in Fermilab's accelerators since its initial commissioning and have typically been overlooked in terms of their criticality and fragility. Vacuum windows allow beam to pass through while creating a boundary between vacuum and air or high vacuum and low vacuum areas. A vacuum window is any relatively thin separation between a volume under vacuum and a volume at atmospheric pressure or vacuum through which primary or secondary beam passes. However, a thin window can also be a thin separation between atmosphere and a pressurized gas.

Vacuum window assemblies must provide reliable mechanical performance to handle both a static differential pressure between the atmosphere and the vacuum, and occasional pressure cycling when the vacuum system is vented for maintenance. However, the window must be thin enough to minimize material irradiation and beam scattering. Figure 1 illustrates an example of how window material can have a dramatic effect on beam scattering. The beam emittance, which is a typical figure-of-merit describing the beam size and angular spread, depends strongly on both the window thickness and material. Therefore, effective window design is a compromise between mechanical reliability, cost, and minimizing the effect on the beam as per the requirements of the beam-line.

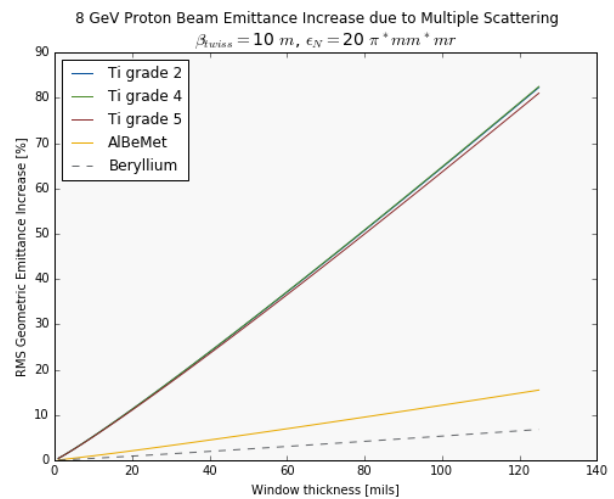

Figure 1: Beam emittance increases as a function of vacuum window thickness (in thousandths of an inch) for different common materials.

Depending on the application and beam requirements, it can be challenging to design a window that meets both mechanical and beam physics constraints.

---
* cader@fnal.gov

## WINDOW DESIGN

A critical aspect of material selection of vacuum windows is examining the costs and benefits of using different materials. The costs of fabricating and operating a new design is up to $92K and most of the cost is the engineering analysis time.

Research and development on vacuum windows is not typically done and there is not a consistent standardized design for the windows. Engineers typically select previously used designs and do not explore other safer or cost-effective designs or materials.

Titanium vacuum windows are the most common and fabricated by an electron-beam process which involves first sandwiching the foil between two Titanium weldments/rings (Figure 2). This sub-assembly is leak checked and then hand welded into a custom titanium conflat.

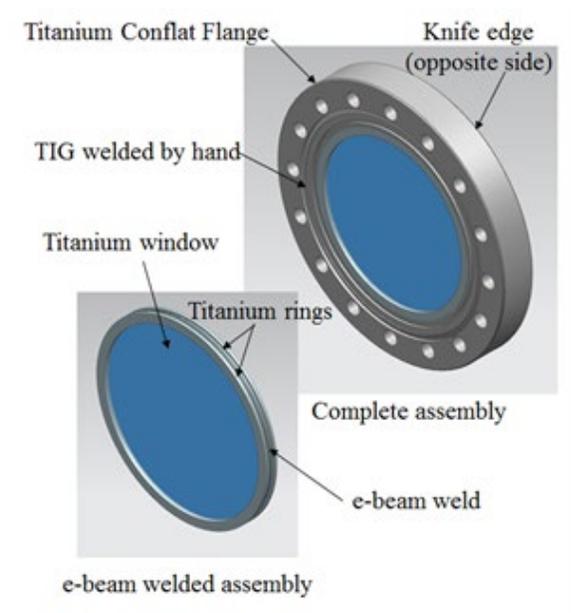

Figure 2: Titanium vacuum window assembly.

A new rectangular shape is now being investigated and it consists of a 6-inch by 3-inch rectangular grade 5 titanium foil, .004" thick. This foil is placed between two support rings made of grade 2 titanium. These support rings have a .05" radius to relieve stress when the foil window flexes under vacuum. Figure 3 below shows how the support rings and foil window attach.

The electron beam welded assembly is then placed into an 8" diameter grade 2 titanium conflat flange that is prepared to receive the welded assembly and welded into place (Figure 3).

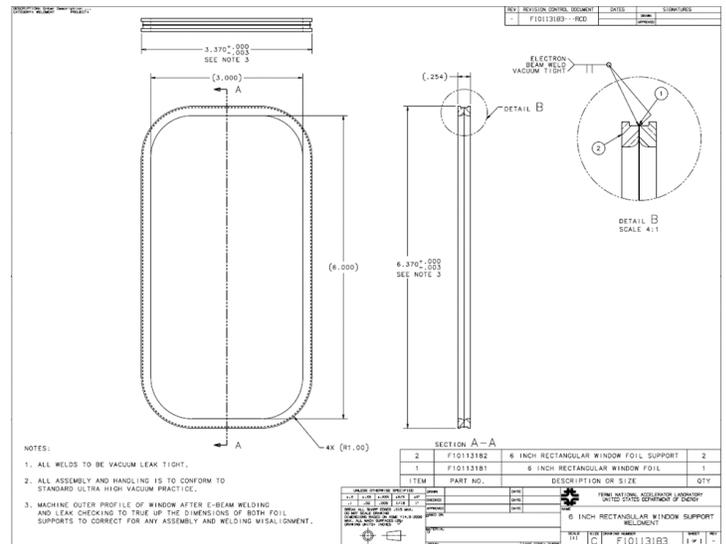

Figure 3: Rectangular window support rings and foil.

## MATERIAL AND FABRICATION

When designing a vacuum window, one must be sure that the thickness remains small or that the material has a small atomic number to reduce interaction with the beam. However, the window cannot be arbitrarily thin, as it may not be able to withstand the pressure difference between the two faces. Additionally, the heat produced by interaction with the passing beam must be transported away as quickly as possible to the edge; otherwise, the window may be damaged if too much heat accumulates in its center. Another important consideration in vacuum window design is the residual radiation created due to interaction to the beam, both as secondary radiation during beam operation and residual activation over long periods of beam exposure.

Different materials produce different mixes of isotopes (i.e. "transmutation") when activated by charged particle beams, which can have a dramatic effect on the residual radiation of the window; for example, activated copper produces a small amount of cobalt-60, which has both high-energy radiated daughter particles and a long half-life, meaning the copper will emit high dose over a long period of time.

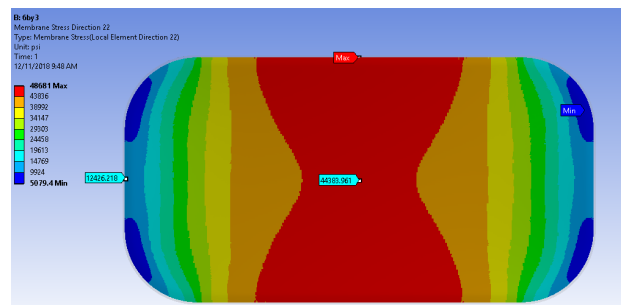

Figure 1: Example of vacuum windows in SY120 (picture taken in 03/2005).

Beryllium has many desirable properties but clean-up costs can be significantly high. There was a beryllium window which failed upstream of the MI-40 abort near the Lambertson and the clean-up was not finished because of the high radiation levels in the area. The decision was made to leave the beryllium in the beam-line until a future date when the abort system is de-commissioned.

Further exploration into recent advancements in fabrication techniques, such as additive manufacturing and ultrasonic welding, could prove beneficial to developing vacuum windows that are both stronger and more transparent to beam than traditional windows.

It is important to fully understand the possible failure modes associated with a vacuum window. A failure root cause analysis (FRCA) is used to help minimize or mitigate risk of potentially harmful effects associated with a vacuum window failure. This FRCA is used to ensure that we build a safe and reliable window and have considered all potential failure modes.

Two primary modes of failure exist: static structural and fatigue. The static structural failure occurs when there are no cyclic forces. These failures typically occur when there is accidental damage of the window during installation. These can be mitigated by a safety procedure. A static loading can cause rupture when the vacuum window stresses are too low or too high. These scenarios are possible when the material becomes radiation damaged and the base material changes. Additionally, the material can also sputter away in some cases as well as corrode due to the environment, which can cause a window failure.

Fatigue can cause a window failure due to cyclic structural or thermal-structural conditions. For instance, when the window is "let-up" to atmosphere from vacuum, the stresses on the window tend to zero, but after so many "let-up" and "pump-down" cycles the window can rupture and no longer hold vacuum. In this specific case, it is imperative to understand the fatigue life of the materials. Additionally, as the beam passes through the window, it will deposit some of its energy into it. That heating and cooling cycle after each pulse of the beam can weaken the material as well. The FRCA chart is intended to guide in the selection and design of a vacuum window.

## BEAM TEST

The goal for the beam test is to expose selected vacuum windows to 120 GeV and 8 GeV protons and observe the effects both in-situ and in a laboratory setting post-exposure. Both beam tests can be accomplished simultaneously in the shared Main Injector and Recycler Ring abort beam-line, which already consists of an air gap of sufficient size for such tests. Similarly, the machine protection system automatically diverts beam to the abort in the event of a drop in the beam permit.

Based on logged data from the past year of running, the absorber has received approximately 1.2E17 protons at 120 GeV and 7.3E16 protons at 8 GeV in the past year of running. This is an attractive location for vacuum window tests because this beam does not take away from the experimental program. Nearby wire scanners and toroids would allow measurements the beam profile and intensity, which will be used to compare the thermal-structural simulations. Additionally, thermal imaging cameras will be used to view the peak temperature of the vacuum windows.

Prior to installing the vacuum windows in the 120 GeV beam-line and the 8 GeV beam-line, the vacuum windows will undergo strain measurements upon initial pump-down. The goal is to effectively measure the strains on the window to correlate the structural model to the physical device. Furthermore, we can cross reference the window estimated deflection using a non-contact device, which has a volumetric accuracy of +-.0013in.

Using a vacuum chamber like the SEM vacuum chamber would allow for testing of two unique windows simultaneously. The design for this vacuum chamber has already been completed and all the vacuum hardware has vendors associated with them

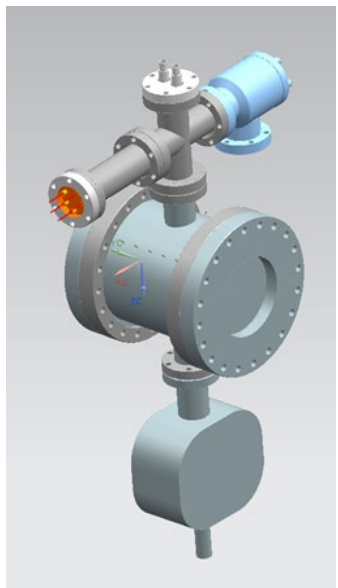

Figure 9: SEM for g-2 Experiment

## CONCLUSION

When the feasibility of a new window design is demonstrated, further irradiation studies in hot cells which is typically used for target materials will be done.

## ACKNOWLEDGEMENTS

This manuscript has been authored by Fermi Research Alliance, LLC under Contract No. DE-AC02-07CH11359 with the U.S. Department of Energy, Office of Science, Office of High Energy Physics.